\documentclass[a4paper,english,12pt]{article}

\usepackage[latin1]{inputenc} 
\usepackage[T1]{fontenc}
\usepackage[safe]{textcomp}
\usepackage{lmodern} 
\usepackage{epsfig}
\usepackage{float}
\usepackage{amstext}  
\usepackage{amsfonts}
\usepackage{amssymb} 
\usepackage{amsmath}
\usepackage{amsthm}
\usepackage{appendix}
\usepackage{bm}   
\usepackage{listings}
\usepackage{enumerate}     
\usepackage[bottom]{footmisc} 
\usepackage{array}           
\usepackage{parskip}        
\usepackage{xcolor} 
\usepackage{color}
\usepackage{framed}
\usepackage[hyphens]{url} 
\usepackage[Sonny]{fncychap}
\usepackage{fancyhdr}
\usepackage{subfig}
\usepackage[raggedright]{sidecap}
\usepackage[plainpages=false]{hyperref}
\usepackage{cite}
\usepackage{slashed}
\usepackage{tikz}
\usepackage[absolute]{textpos}
\usepackage{pstricks}
\usepackage{axodraw4j}
\usepackage{graphicx}
\usepackage{dsfont}
\usepackage{multirow}
\usepackage{booktabs}
\makeindex

\newcommand{\p}{\partial}

\newcommand{\f}[2]{\frac{#1}{#2}}
\newcommand{\sss}[1]{\scriptscriptstyle#1}
\newcommand{\vv}[2]{\left( \begin{array}{c} #1 \\ #2  \end{array} \right)}
\newcommand{\bea}{\begin{eqnarray}}
\newcommand{\eea}{\end{eqnarray}}
\newcommand{\be}{\begin{equation}}
\newcommand{\ee}{\end{equation}}
\newcommand{\ba}{\begin{align}}
\newcommand{\ea}{\end{align}}
\newcommand{\beas}{\begin{eqnarray*}}
\newcommand{\eeas}{\end{eqnarray*}}
\newcommand{\bes}{\begin{equation*}}
\newcommand{\ees}{\end{equation*}}
\newcommand{\bas}{\begin{align*}}
\newcommand{\eas}{\end{align*}}

\newcommand{\ssL}{{\mathcal L}} 
\newcommand{\eps}{{\varepsilon}}
\newcommand{\cd}{{\cdot}}

\newcommand{\Nf}{n_{\scriptscriptstyle{f}}}
\newcommand{\gs}{g_{\scriptscriptstyle{s}}}
\newcommand{\yt}{y_{\scriptscriptstyle{t}}}

\newcommand{\als}{\alpha_{\scriptscriptstyle{s}}}

\newcommand{\lb}{\left(}
\newcommand{\rb}{\right)}

\definecolor{bluemar}{rgb}{0,0,.5}
\definecolor{redmar}{rgb}{.8,0,0}
\definecolor{greenmar}{rgb}{0,.5,0}

\newcommand{\rd}{\color{redmar}}

\newcommand{\bk}{\color{black}}

\newcommand{\ice}[1]{\relax}
\graphicspath{{./}{./figures/}}

\ChTitleVar{\Large\rm\bf}
\ChNameVar{\Large\rm\bf}

\begin{document}

\pagestyle{plain}

\setcounter{tocdepth}{1}
\pagenumbering{arabic}
\setlength{\fboxrule}{0.5 mm} 

\begin{flushright}
TTP12-036\\
SFB/CPP-12-69
\par\end{flushright}
\vskip 0.2cm
\begin{center}
{\huge\bf Vacuum stability in the SM and the three-loop $\beta$-function for the Higgs self-interaction}
\vskip 0.4cm
M. F. Zoller\footnote{email: max.zoller@kit.edu}\\[1ex]
{Institut f\"ur Theoretische Teilchenphysik, Karlsruhe
  Institute of Technology (KIT),\\ D-76128 Karlsruhe, Germany}
\vskip 0.5cm
{\bf Abstract}\\[1ex]
\end{center}
In this article the stability of the Standard Model (SM) vacuum
in the presence of radiative corrections and for a Higgs boson with a mass in the vicinity of $125$ GeV is discussed.
The central piece in this discussion will be the Higgs self-interaction $\lambda$ and
its evolution with the energy scale of a given physical process.
This is described by the $\beta$-function to which we recently computed analytically the
dominant three-loop contributions \cite{Chetyrkin:2012rz}.\footnote{In
\cite{Chetyrkin:2012rz} we also give the dominant contributions to the
$\beta$-functions for the top-Yukawa coupling, the strong coupling and
the anomalous dimensions of the scalar, gluon and quark fields 
in the unbroken phase of the Standard Model at three-loop level.}
These are mainly the QCD and top-Yukawa corrections as well as
the contributions from the Higgs self-interaction itself. We will see that
for a Higgs boson with a mass of about $125$ GeV the question whether the SM vacuum is stable
and therefore whether the SM could be
valid up to Planck scale cannot be answered with certainty due to large
experimental uncertainties, mainly in the top quark mass.

\section{The Higgs potential and the stability of the SM vacuum}
\begin{figure}[h!]\centering
 \includegraphics[width=0.5\linewidth]{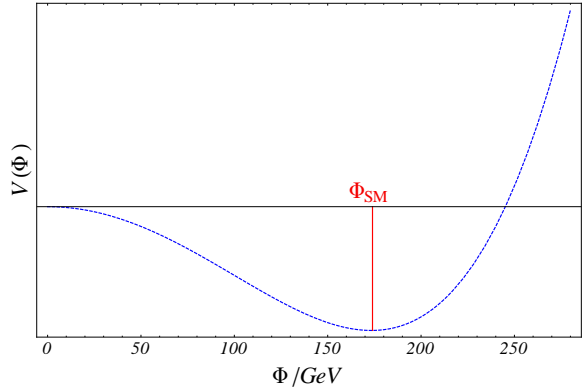}
\caption{The SM Higgs potential at tree-level for $M_H=125$ GeV}
\label{Vtree_125}
\end{figure}

In the SM the Higgs potential at tree-level appears as part of the Lagrangian for a scalar SU($2$)-doublet field:
\be \ssL_{\sss{\Phi}}=\p_\mu \Phi^\dagger \p^\mu \Phi-
\underbrace{\left(m^2\Phi^\dagger\Phi+\lambda \lb\Phi^\dagger\Phi\rb^2\right)}_{V(\Phi)}, \qquad
\Phi=\vv{\Phi_1}{\Phi_2}. \label{Vphidef} \ee
This doublet aquires a non-zero vacuum expectation value (VEV) under spontaneous symmetry breaking (SSB) and
we get the Higgs field, three Goldstone bosons\footnote{These Goldstone bosons can be absorbed
by the massive W and Z bosons.} and the masses of the SM particles:
\be \Phi=\vv{\Phi_1}{\Phi_2}\xrightarrow{SSB} \vv{\Phi^+}{\f{1}{\sqrt{2}}(v+H+i\chi)},
\qquad|\Phi_{SM}|=\sqrt{\f{-m^2}{2\lambda}}=\f{v}{\sqrt{2}}\ee
At tree level the mass of the Higgs boson is then given by \be M_H^2=-2m^2=2\lambda v^2. \ee For $M_H=125$ GeV the Higgs potential
is shown in Fig. \ref{Vtree_125}.
If we now include radiative corrections we have to consider the effective potential 
$V_{eff}(\lambda(\Lambda),g_i(\Lambda),\Phi(\Lambda))$
as introduced in \cite{PhysRevD.7.1888}. All couplings and fields undergo an evolution up to some scale
$\Lambda$ where the theory ceases to be valid.\footnote{With $t=\log\lb\f{\Lambda}{\mu_0}\rb$ we have the field
$\Phi(\Lambda)=\Phi_{cl}\cd \text{exp} \lb \int\limits_0^t \!dt' \gamma_\Phi(\lambda(t'),g_i(t'))dt'\rb$ 
where $\Phi_{cl}$ is the classical field in the absence of radiative corrections.
$\mu_0$ is the scale at which we start the running of the paramters.}
Here we want to investigate whether a scenario in which the SM provides a
good description of nature up to the Planck scale, i.e. \mbox{$\Lambda\sim 10^{18}$ GeV}, is possible.\footnote{An extended
model is expected to be needed at this scale due to gravity.} 
\begin{figure}[h!]\begin{center}
                   \begin{minipage}{0.45\linewidth}
                    \includegraphics[width=\linewidth]{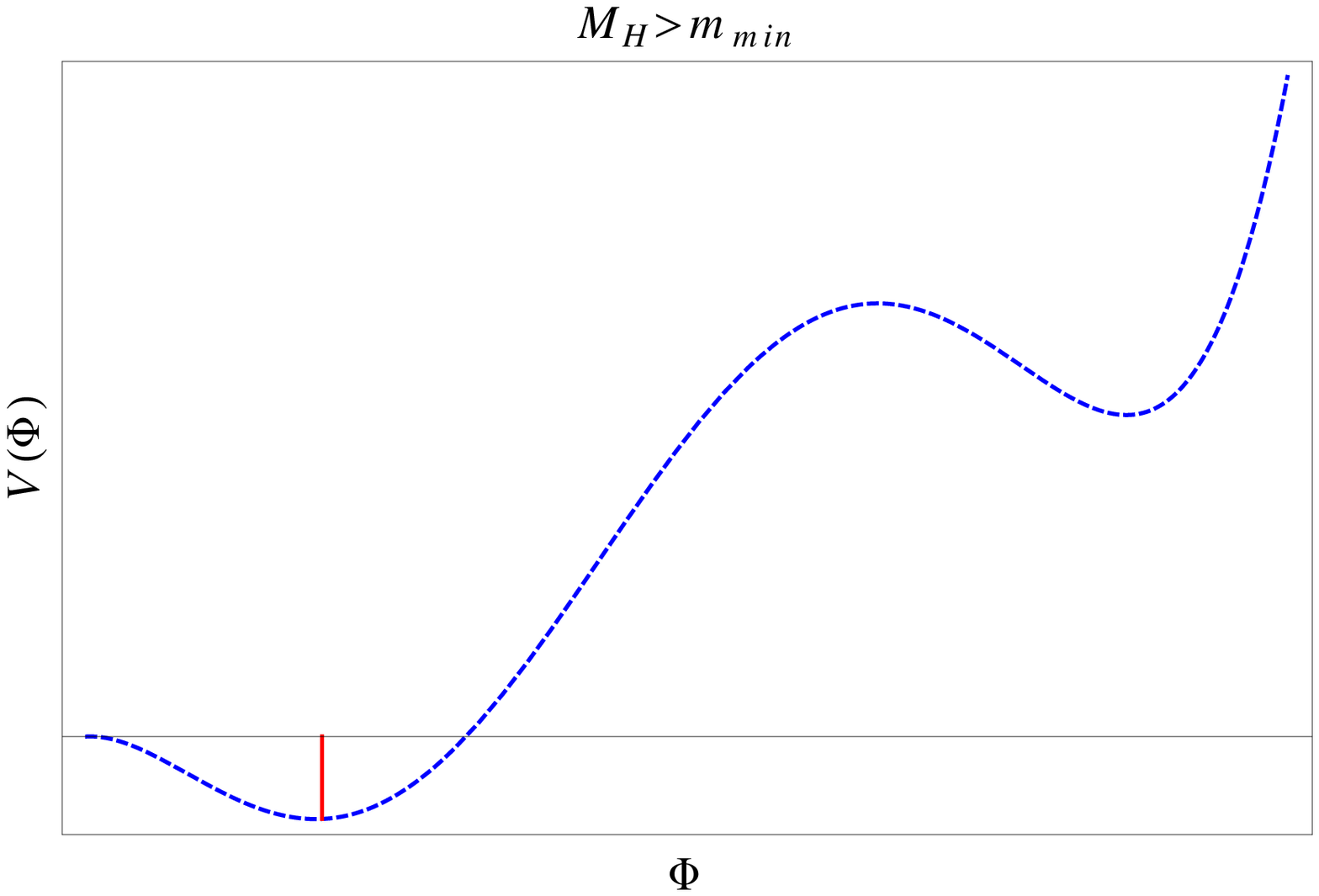}
                   \end{minipage}
                   \begin{minipage}{0.45\linewidth}
                    \includegraphics[width=\linewidth]{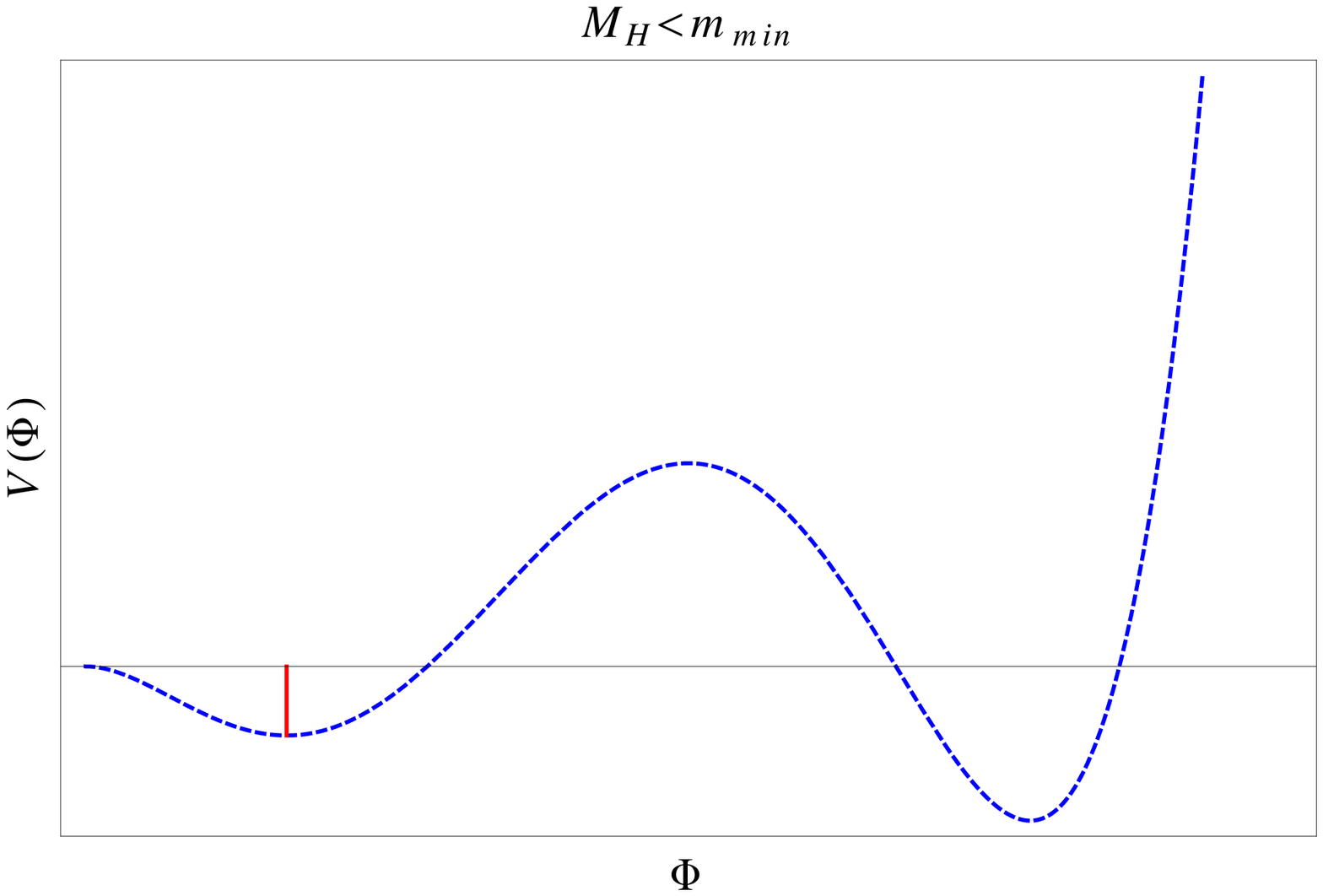}
                   \end{minipage}
                  \end{center}
\caption{The effective Higgs potential}
\label{Veffective}
\end{figure}
\normalsize
The general shape of this effective potential for the SM is shown in Fig. \ref{Veffective} for the cases
of a Higgs mass larger and smaller than a critical value $m_{min}$, the minimal stability bound\footnote{There
is also an upper bound $m_{max}$ on the Higgs mass
from the requirement that no Landau pole appears at energies $\mu\leq \Lambda$.} (see also \cite{Bezrukov:2012sa}).
If the Higgs mass is chosen below this critical value a second minimum develops which is lower than the SM one.
This implies that the SM vacuum is no longer stable, i.e. it can tunnel into this energetically favoured state.
Since this is in contradiction to our observation\footnote{It is however not possible to exclude
that we live in metastable universe, i.e. the lifetime of the (local) SM minimum could be longer than the age of the universe.}
we are led to the conclusion
that our theory is incomplete and that new physics has to enter between Fermi and Planck scale.
It has been demonstrated in \cite{Altarelli1994141} that for $\Phi \sim \Lambda \gg \mu_0$ a good approximation
for the effective potential is
\be V_{eff}[\Phi]\approx \lambda(\Lambda) \Phi^4(\Lambda)+{\cal{O}}(\lambda^2(\Lambda),g_i^2(\Lambda)), \ee
\normalsize
which means that the stability of the SM vacuum is approximately
equivalent to the question whether $\lambda$ stays positive up to the scale $\Lambda$ (see also \cite{Cabibbo:1979ay,Ford:1992mv}).
\begin{figure}[h!]\begin{center}
                    \includegraphics[width=0.8\linewidth]{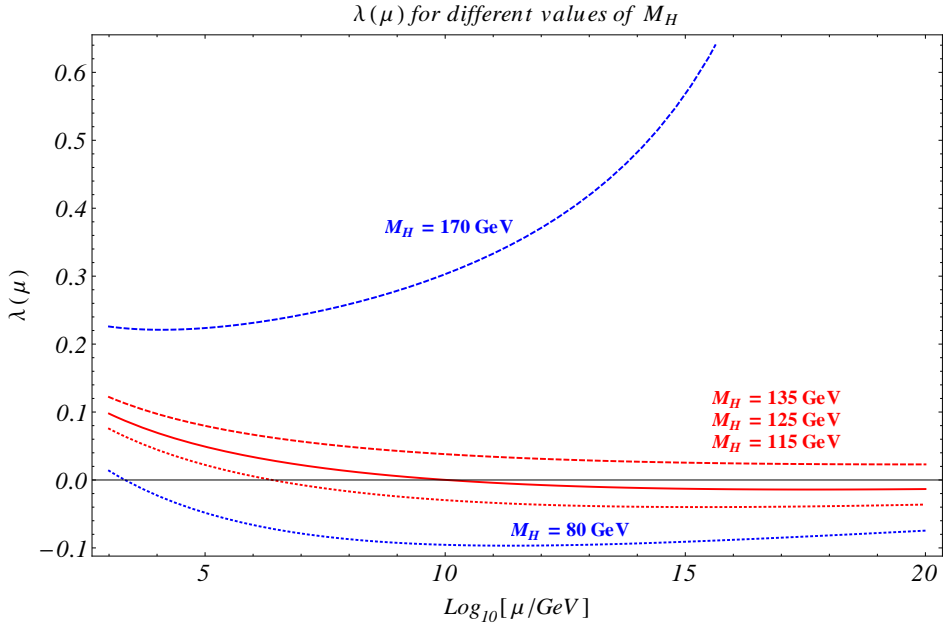}
                  \end{center}
\caption{The behaviour of $\lambda(\mu)$ for different Higgs mass values}
\label{lambdarunning}
\end{figure}
Fig. \ref{lambdarunning} shows the evolution of $\lambda$ for different Higgs mass values. For a large enough Higgs mass
$\lambda$ diverges quickly\footnote{We find a Landau pole below the Planck scale
for $M_H>m_{max} \approx 175$ GeV (see also \cite{Maiani:1977cg,Cabibbo:1979ay,Lindner:1985uk,Hambye:1996wb}).} 
and for a small Higgs mass $\lambda$ would become negative at a relatively low scale.\footnote{Eventually $\lambda$ will increase again
and reach a Landau pole as well due to the the evolution of the gauge couplings and $\yt$. At large scales (above $10^{16}$ GeV) the electroweak couplings, especially the $U(1)$ coupling $g_1$, start
to be the dominant contributions. In contrast $\yt$, which is responsible for the decrease of $\lambda$ at lower scales, becomes small.
In order to see this one has to extrapolate the evolution of $\lambda$ well beyond the Planck scale.
This is also the reason why the effective potential stays bounded from below at large field strengths $\Phi$ beyond the second minimum.}
The interesting region is around $125$ GeV where $\lambda$ is very close to zero at the Planck scale!
As the minimal stability bound has been estimated to be about $129 \pm 3$ GeV 
\cite{Bezrukov:2012sa,EliasMiro:2011aa}, which is very close 
to the mass of the boson recently descovered  at the LHC, the question of vacuum stability in the SM becomes one for
precision physics.
This serves as a strong motivation for calculating the three-loop $\beta$-function of the Higgs self-interaction
which describes the evolution of this crucial coupling as well as the $\beta$-functions of the relevant SM parameters
on which $\beta_\lambda$ depends.

\section{The  three-loop $\beta$-function for the Higgs self-interaction}
The $\beta$-function is defined as the derivative of the coupling with respect to the renormalization scale:
\be \beta_{\lambda}(\lambda,\yt,g_i,\ldots)=\mu^2\f{d}{d\mu^2} \lambda(\mu).\ee
This object has been known at the two-loop level for a while including the dependence on
all the gauge couplings $g_1$,$g_2$ and $g_s$, 
the quartic Higgs self-interaction $\lambda$ and the Yukawa couplings which give mass to the fermions 
\cite{Fischler1982385,Machacek1984221,Ford:1992pn,2loopbetayukawa}.
The one-loop and two-loop results for all SM couplings have been known for a long time
\cite{PhysRevLett.30.1343,PhysRevLett.30.1346,Jones1974531,Tarasov:1976ef,PhysRevLett.33.244,
Egorian:1978zx,PhysRevD.25.581,Fischler:1981is,Fischler1982385,Jack1985472,
Machacek198383,Machacek1984221,Machacek198570,2loopbetayukawa,Ford:1992pn} as have been partial three-loop
results \cite{Curtright:1979mg,Jones:1980fx,Tarasov:1980au,3loopbetaqcd,Steinhauser:1998cm,Pickering:2001aq}.
Four-loop $\beta$-functions are available for QCD \cite{4loopbetaqcd,Czakon:2004bu} and 
the purely scalar part of the SM \cite{Brezin:1974xi,Brezin:1973,Kazakov_betalambda}.

As there are many Feynman diagrams at the three-loop level and the treatment of $\gamma_5$ matrices in dimensional regularization
poses a serious problem
at this order for the Higgs and Yukawa sector we try to find the dominant contributions to the evolution of the Higgs self-interaction
at the scales that we are intersted in. At the scale of the top mass \mbox{$\mu=M_t=172.9$ GeV} we find
the strong coupling \mbox{$\gs \approx 1.17$}  and the
top-Yukawa coupling \mbox{$\yt \approx 0.93$}  to be
much larger than the electroweak couplings \mbox{$g_2 \approx 0.65$} and \mbox{$g_1 \approx 0.36$.} The Higgs self-interaction
for a Higgs boson around $125$ GeV is \mbox{$\lambda(M_{\sss{H}})\approx 0.13$.}
\footnote{
These couplings enter at every loop order with characteristic factors
$\f{\gs^2}{4\pi}\approx 0.11$, $\f{g_2^2}{4\pi}\approx 0.03$,
$\f{g_1^2}{4\pi}\approx 0.01$ and $\f{\yt^2}{4\pi}\approx 0.07$. The Higgs self-coupling is already present at tree level and
enters at each order linearly as $\f{\lambda}{4\pi}\approx 0.01$.
The second-largest Yukawa coupling to be considered would be
\mbox{$y_b=\sqrt{2}\f{M_b}{v}\approx 0.02$} which is negligible in comparison.}
From this we conclude that a simplified model containing only $\gs$, $\yt$ and $\lambda$ will give the numerically largest
terms for $\beta_\lambda$ at three-loop level.  When combinen with the full SM result at one and two-loop level this leads to
 the evolution of $\lambda$ to the highest precision so far.
The Lagrangian of our model consists of three pieces:
\be
\ssL=\ssL_{\sss{QCD}}+\ssL_{\sss{\Phi}}+\ssL_{\sss{\yt}}.
\ee
with the standard QCD Lagrangian, the Higgs part as defined in \eqref{Vphidef} and the top-Yukawa sector
\be
\begin{split}
\ssL_{\sss{\yt}}
&=-\yt \left\{ 
\bar{t}_{\sss{R}}\lb \Phi_2, -\Phi_1\rb\cd \vv{t}{b}_{\sss{L}}
+\,\lb \bar{t}, \bar{b}\rb_{\sss{L}}\cd \vv{\Phi^{*}_2}{-\Phi^{*}_1}\,t_{\sss{R}}
\right\}\\
\end{split} 
\label{LYuk} 
\ee
The indices L and R indicate the left- and right-handed part of the Dirac fermion fields as obtained by the projectors
\be 
P_{\sss{L}}=\f{1}{2}\lb 1-\gamma_5\rb \qquad P_{\sss{R}}=\f{1}{2}\lb 1+\gamma_5\rb
{}.
\ee
Hence $\gamma_5$ enters into our calculation and has to be treated carefully as described below.
\subsection{Calculation \label{calc}}
In order to obtain the $\beta$-functions for a coupling, here $\lambda$ or $\yt$,
we have to compute the renormalization constant for a vertex involving this coupling and the field renormalization constants
for the external legs of this vertex. The latter is done by computing loop corrections to the respective propagators.
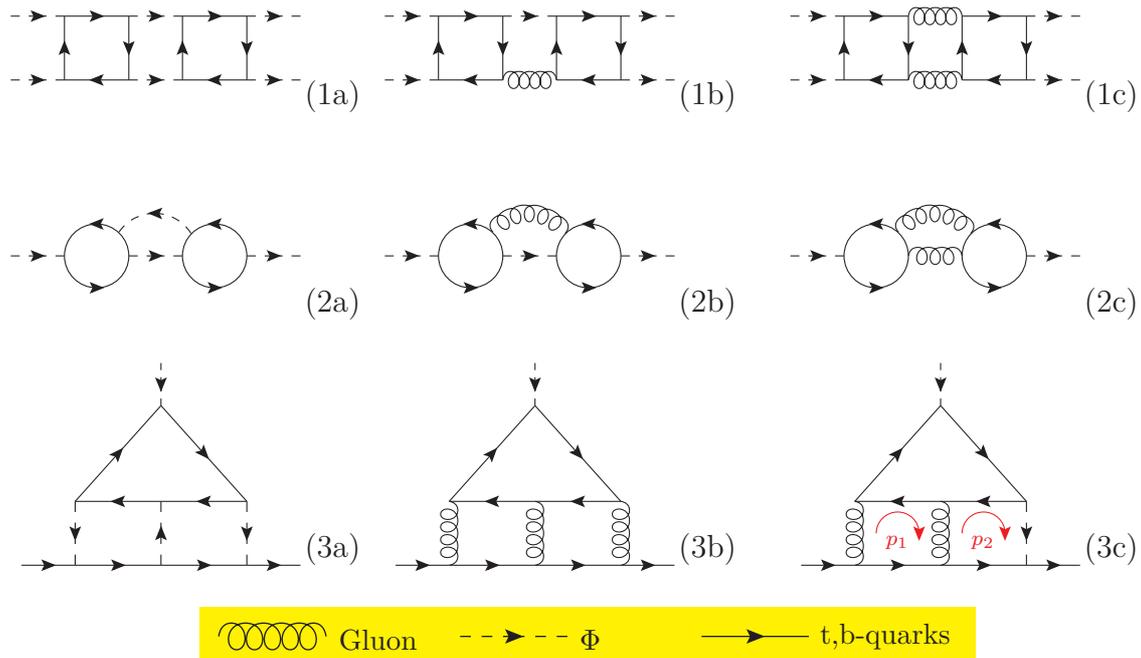
\begin{figure}[h!] \vspace{-20mm} \begin{center}
 \begin{align*}
    \begin{picture}(140,100) (0,0) \SetScale{0.8}
    \SetColor{Black}
    \DashArrowLine(0,15)(25,15){4}
    \DashArrowLine(0,45)(25,45){4}
    \DashArrowLine(110,15)(135,15){4}
    \DashArrowLine(110,45)(135,45){4}
    \ArrowLine(25,45)(55,45)
    \ArrowLine(55,15)(25,15)
    \ArrowLine(25,15)(25,45)
    \ArrowLine(55,45)(55,15)
    \DashArrowLine(55,15)(80,15){4}
    \DashArrowLine(55,45)(80,45){4}
    \ArrowLine(80,45)(110,45)
    \ArrowLine(110,15)(80,15)
    \ArrowLine(80,15)(80,45)
    \ArrowLine(110,45)(110,15)
    \Text(110,0)[lb]{\Black{(1a)}}
  \end{picture} 
&    \begin{picture}(140,100) (0,0) \SetScale{0.8}
    \SetColor{Black}
    \DashArrowLine(0,15)(25,15){4}
    \DashArrowLine(0,45)(25,45){4}
    \DashArrowLine(110,15)(135,15){4}
    \DashArrowLine(110,45)(135,45){4}
    \ArrowLine(25,45)(55,45)
    \ArrowLine(55,15)(25,15)
    \ArrowLine(25,15)(25,45)
    \ArrowLine(55,45)(55,15)
    \Gluon(55,15)(80,15){4}{4}
    \DashArrowLine(55,45)(80,45){4}
    \ArrowLine(80,45)(110,45)
    \ArrowLine(110,15)(80,15)
    \ArrowLine(80,15)(80,45)
    \ArrowLine(110,45)(110,15)
    \Text(110,0)[lb]{\Black{(1b)}}
  \end{picture} 
&    \begin{picture}(140,100) (0,0) \SetScale{0.8}
    \SetColor{Black}
    \DashArrowLine(0,15)(25,15){4}
    \DashArrowLine(0,45)(25,45){4}
    \DashArrowLine(110,15)(135,15){4}
    \DashArrowLine(110,45)(135,45){4}
    \ArrowLine(25,45)(55,45)
    \ArrowLine(55,15)(25,15)
    \ArrowLine(25,15)(25,45)
    \ArrowLine(55,45)(55,15)
    \Gluon(55,15)(80,15){4}{4}
    \Gluon(55,45)(80,45){4}{4}
    \ArrowLine(80,45)(110,45)
    \ArrowLine(110,15)(80,15)
    \ArrowLine(80,15)(80,45)
    \ArrowLine(110,45)(110,15)
    \Text(110,0)[lb]{\Black{(1c)}}
  \end{picture}\\[-10mm]
  \begin{picture}(140,100) (0,0)\SetScale{0.8}
    \SetColor{Black}
    \DashArrowLine(0,30)(25,30){4}
    \DashArrowLine(110,30)(135,30){4}
    \ArrowArc(40,30)(15,180,0)
    \ArrowArc(40,30)(15,0,180)
    \ArrowArc(95,30)(15,180,0)
    \ArrowArc(95,30)(15,0,180)
    \DashArrowLine(55,30)(80,30){4}
    \DashArrowArc(67,30)(20,32,146){4}
    \Text(110,0)[lb]{\Black{(2a)}}
  \end{picture} 
&    \begin{picture}(140,100) (0,0)\SetScale{0.8}
    \SetColor{Black}
    \DashArrowLine(0,30)(25,30){4}
    \DashArrowLine(110,30)(135,30){4}
    \ArrowArc(40,30)(15,180,0)
    \ArrowArc(40,30)(15,0,180)
    \ArrowArc(95,30)(15,180,0)
    \ArrowArc(95,30)(15,0,180)
    \DashArrowLine(55,30)(80,30){4}
    \GlueArc(67,30)(20,32,146){4}{5}
    \Text(110,0)[lb]{\Black{(2b)}}
  \end{picture} 
&  \begin{picture}(140,100) (0,0)\SetScale{0.8}
    \SetColor{Black}
    \DashArrowLine(0,30)(25,30){4}
    \DashArrowLine(110,30)(135,30){4}
    \ArrowArc(40,30)(15,180,0)
    \ArrowArc(40,30)(15,0,180)
    \ArrowArc(95,30)(15,180,0)
    \ArrowArc(95,30)(15,0,180)
    \Gluon(55,30)(80,30){4}{3}
    \GlueArc(67,30)(20,32,146){4}{5}
    \Text(110,0)[lb]{\Black{(2c)}}
  \end{picture}\\[-5mm]
    \begin{picture}(140,100) (0,0)\SetScale{0.8}
    \SetWidth{0.5}
    \SetColor{Black}
    \DashArrowLine(70,95)(70,75){4}
    \ArrowLine(70,75)(110,30)
    \ArrowLine(110,30)(70,30)
    \ArrowLine(70,30)(30,30)
    \ArrowLine(30,30)(70,75)
    \DashArrowLine(30,30)(30,0){4}
    \DashArrowLine(70,0)(70,30){4}
    \DashArrowLine(110,30)(110,0){4}
    \ArrowLine(5,0)(30,0)
    \ArrowLine(30,0)(70,0)
    \ArrowLine(70,0)(110,0)
    \ArrowLine(110,0)(135,0)
    \Text(110,0)[lb]{\Black{(3a)}}
  \end{picture}
&    \begin{picture}(140,100) (0,0)\SetScale{0.8}
    \SetWidth{0.5}
    \SetColor{Black}
    \DashArrowLine(70,95)(70,75){4}
    \ArrowLine(70,75)(110,30)
    \ArrowLine(110,30)(70,30)
    \ArrowLine(70,30)(30,30)
    \ArrowLine(30,30)(70,75)
    \Gluon(30,30)(30,0){4}{4}
    \Gluon(70,30)(70,0){4}{4}
    \Gluon(110,30)(110,0){4}{4}
    \ArrowLine(5,0)(30,0)
    \ArrowLine(30,0)(70,0)
    \ArrowLine(70,0)(110,0)
    \ArrowLine(110,0)(135,0)
    \Text(110,0)[lb]{\Black{(3b)}}
  \end{picture}
&    \begin{picture}(140,100) (0,0)\SetScale{0.8}
    \SetWidth{0.5}
    \SetColor{Black}
    \DashArrowLine(70,95)(70,75){4}
    \ArrowLine(70,75)(110,30)
    \ArrowLine(110,30)(70,30)
    \ArrowLine(70,30)(30,30)
    \ArrowLine(30,30)(70,75)
    \Gluon(30,30)(30,0){4}{4}
    \Gluon(70,30)(70,0){4}{4}
    \DashArrowLine(110,30)(110,0){4}
    \ArrowLine(5,0)(30,0)
    \ArrowLine(30,0)(70,0)
    \ArrowLine(70,0)(110,0)
    \ArrowLine(110,0)(135,0)
    \SetColor{Red}
    \LongArrowArcn(50,15)(10,180,0)
    \LongArrowArcn(90,15)(10,180,0)
    \Text(40,6)[cb]{\scriptsize \Red{$p_1$}}
    \Text(72,6)[cb]{\scriptsize \Red{$p_2$}}
    \SetColor{Black}
    \Text(110,0)[lb]{\Black{(3c)}}
  \end{picture}
\end{align*}
\colorbox{yellow}{
\begin{picture}(280,15) (0,-7)
\Gluon(0,0)(40,0){5}{5}  \Text(45,-5)[lb]{\small{\Black{Gluon}}}
\DashArrowLine(90,0)(130,0){5}  \Text(135,-5)[lb]{\small{\Black{$\Phi$}}}
\ArrowLine(180,0)(220,0)  \Text(225,-5)[lb]{\small{\Black{t,b-quarks}}}
\end{picture}} \end{center} \caption{Sample Feynman diagrams for the calculation of renormalization constants for
the quartic Higgs vertex, the Higgs self-energy and a Yukawa vertex at three-loop order.}
\label{Feynmandias}
\end{figure}
The renormalization constant for $\yt$ for example can be computed as\footnote{Note that in general left and right-handed fields
have to be renormalized separately.}
\be 
Z_{\yt}=\f{Z^{\text{(Yukawa vertex)}}}{\sqrt{Z^{\text{(top field)}}_{L}Z^{\text{(top field)}}_{R}Z^{\text{(Higgs field)}}}}.
\ee
The $\beta$-function $\beta_{\yt}$ can then be computed from the requirement that the bare Yukawa coupling
\be \yt^{bare}=Z_{\yt}[g_i(\mu),\yt(\mu),\lambda(\mu)]\, \yt(\mu) \label{betaren} \ee
is independent of the renormalization scale $\mu$ and hence its $\mu^2$-derivative 
must vanish.

For the purpose of calculating renormalization constants and $\beta$-functions we only need the UV-divergent part of all these diagrams.
Nevertheless, there are two issues which have to be considered carefully in a computation of this type.
The first problem is the complicated nature of our integrals when we have arbitrary momenta flowing into the external legs of 
our Feynman diagrams. A nice feature of the $\overline{\text{MS}}$-scheme however is that renormalization constants do not depend on
external momenta\footnote{To be precise, UV-divergent terms can only depend polynomially on external momenta (and masses).}, so
we can set those to zero. Unfortunately, this introduces artificial IR-divergences into our diagrams which cannot
be distinguished from the UV ones in dimensional regularization. In many cases this problem can be avoided by 
setting all external momenta to zero except for one which enters at one leg and exits at another. If we have no masses, as we do here,
the propagator-like integrals resulting from this method can be computed with the FORM
package MINCER \cite{MINCER} up to three-loop order.
For the $\lambda$-vertex and its radiative corrections this fails, however, as there are still IR-divergent diagrams.
A method to compute only the UV-divergences without having to worry about the IR ones has been described
in \cite{Misiak:1994zw, beta_den_comp}. The trick is to introduce the same auxiliary mass into every propagator denominator,
Taylor expand in all the external momenta and introduce all possible counterterms for the auxiliary mass in order to cancel
subdivergences arising from this new mass. The resulting massive tadpole integrals can be computed with the FORM package MATAD \cite{MATAD}.
This method will yield the correct UV-pole part of the calculated Feynman diagrams
(but not the correct finite part) which is enough for the computation of $\beta$-functions. Where possible both methods have been used for the
calculation in order to have an independent check.

The second problem is the treatment of $\gamma_5$ matrices, which arise here from the projectors in the Yukawa sector, in $d$
space-time dimensions. This matrix is only well-defined in four dimensions:
\be \gamma_5=i\gamma^0\gamma^1\gamma^2\gamma^3=\f{i}{4!}\eps_{\mu\nu\rho\sigma} \gamma^\mu\gamma^\nu\gamma^\rho\gamma^\sigma \,\text{ where }
 \,\eps_{0123}=1=-\eps^{0123} \label{gamma5} {}.
\ee
A naive treatment of $\gamma_5$, i.e. using the relations $\{\gamma_5,\gamma_\mu\}=0$ and $\gamma_5^2=\mathds{1}$ to eliminate 
as many $\gamma_5$ as possible and then discarding all
terms which still have one $\gamma_5$ in them, can only be applied to external fermion lines and closed fermion loops
with less than four Lorentz indices or momenta flowing in or out of the fermion loop in question.\footnote{Four
different Lorentz indices are needed to support a non-vanishing $\eps_{\mu\nu\rho\sigma}$ after the trace
over the fermion line has been performed. These can be the Lorentz indices of gauge boson vertices or they can be contracted
with internal momenta from other loops. External momenta can be set to zero without changing the UV-divergent part and can hence be ignored in this consideration.}
An example for a problematic diagram is shown in Fig. \ref{Feynmandias} (3c). The Lorentz indices of the two gluons 
connected to the closed fermion loop and the two loop momenta $p_1$ and $p_2$ offer the possibility of a 
non-trivial $\gamma_5$ contribution from such a diagram. And indeed, if we apply the treatment suggested by 't Hooft and Veltman in 
\cite{'tHooft:1972fi}, i.e. using the definition with $\eps_{\mu\nu\rho\sigma}$ from \eqref{gamma5} and
contracting the $\eps$-tensor from the closed fermion loop with the one from the external fermion line or a projector $\propto \gamma_5$ 
acting on this external line to make the integral scalar, we find a sizable contribution. But even this treatment is not exact, in fact
it is only correct up to an error of $\mathcal{O}(\eps)$. Fortunately, there are only first order poles in $\eps$ from these integrals
which makes the UV-divergent part and therefore our renormalization constants correct. Once again, the finite part calculated for
these diagrams is unreliable but luckily not needed here.

\subsection{Result}
The result for the $\beta$-function of the Higgs self-interaction 
\be \mu^2\f{d}{d\mu^2} \lambda(\mu)=\beta_{\sss{\lambda}}(\gs,\yt,\lambda)
=\sum \limits_{n=1}^{\infty} \f{1}{(16\pi^2)^{n}}\,\beta_{\sss{\lambda}}^{(n)}(\gs,\yt,\lambda) \ee
in our simplified version of the Standard Model is given by
\bes \begin{split}
\beta_{\sss{\lambda}}^{(1)}=& 12 \,  \lambda^2         + 6\, \yt^2 \lambda      - 3\,\yt^4              ,\\
\beta_{\sss{\lambda}}^{(2)}=&         - 156\,\lambda^3            
       - 72\, \yt^2 \lambda^2         
       - \f{3}{2}\, \yt^4 \lambda           
       + 15\, \yt^6            
       + 40\, \gs^2 \yt^2 \lambda          
       - 16\, \gs^2 \yt^4              ,\\   
\beta_{\sss{\lambda}}^{(3)}=& 
       \lambda^4   \left(           3588          + 2016 \zeta_{3}          \right)
       + 873\, \yt^2 \lambda^3           
       + \yt^4 \lambda^2   \left(          \f{1719}{2}          + 756 \zeta_{3}          \right)\\
       &+ \yt^6 \lambda   \left(           \f{117}{8}          - 198 \zeta_{3}          \right)
       - \rd \yt^8 \bk  \left(           \f{1599}{8}          + 36 \zeta_{3}          \right)\\
       &+ \gs^2 \yt^2 \lambda^2   \left(          - 1224          + 1152 \zeta_{3}          \right)
       + \rd\gs^2 \yt^4 \lambda \bk  \left(           895          - 1296 \zeta_{3}          \right)\\
       &+ \rd\gs^2 \yt^6\bk   \left(          - 38          + 240 \zeta_{3}          \right)
       + \rd\gs^4 \yt^2 \lambda\bk   \left(           \f{1820}{3}          - 32 \Nf          - 48 \zeta_{3}          \right)\\
       &+ \rd\gs^4 \yt^4\bk   \left(          - \f{626}{3}          + 20 \Nf
         + 32 \zeta_{3}          \right)\\
\end{split} \label{beta:la} \ees 
in the $\overline{\text{MS}}$-scheme. $\beta_{\sss{\lambda}}^{(1)}$, $\beta_{\sss{\lambda}}^{(2)}$ and $\beta_{\sss{\lambda}}^{(3)}$
are the one, two and three-loop results respectively.
To get an idea of the size of the individual terms and the overall result
we evaluate $\beta_{\sss{\lambda}}$ at the scale $\mu=M_Z$ 
(with an assumed Higgs mass of \mbox{125 GeV} and the number of fermion flavours \mbox{$\Nf=6$}) which
yields a value of \mbox{$\beta_{\sss{\lambda}} \sim (-0.01)$} at the one-loop level. 
The two and three-loop contributions change this result
by \mbox{$\sim 1\%$} and \mbox{$\sim (-0.04)\%$} respectively which is quite small e.g. in comparison with the
$\beta$-function for the top-Yukawa coupling where we get a value of $\beta_{\yt} \sim(-0.023)$
and corrections of $\sim 16.6\%$ and $\sim 0.7\%$ at two and three-loop level. The full result for $\beta_{\yt}$
can be found in \cite{Chetyrkin:2012rz}.
If we have a look at the the numerically largest individual terms in \eqref{beta:la} we find a curious behaviour:
\be \f{\beta_{\sss{\lambda}}^{(3)}}{(16\pi^2)^3}(\mu=M_Z)=(
\underbrace{+7.9}_{\rd\gs^2\yt^6\bk}\quad
\underbrace{-4.8}_{\rd\yt^8\bk}\quad
\underbrace{-3.1}_{\rd\gs^2\yt^4\lambda\bk}\quad
\underbrace{-2.5}_{\rd\gs^4\yt^4\bk}\quad
\underbrace{+2.6}_{\rd\gs^4\yt^2\lambda\bk})\,\times\,10^{-5}. \ee
There is a strong cancellation between these terms making the overall effect almost two orders
of magnitude smaller than the largest individual contributions.\footnote{Note that for a significantly different Higgs mass
this would not happen to such extent.}
This significantly improves the
convergence of the perturbation series for $\beta_{\lambda}$  and makes the remaining theoretical uncertainty small. 
\section{The evolution of the Higgs self-coupling \label{la:evolution}}
Now we want to investigate the
effect of the new three-loop result $\beta_{\sss{\lambda}}^{(3)}$ on the
running of $\lambda$ and therefore its effect on the stability of the
electroweak vacuum in the SM.
\begin{figure}[h!]
\includegraphics[width=0.95\textwidth]{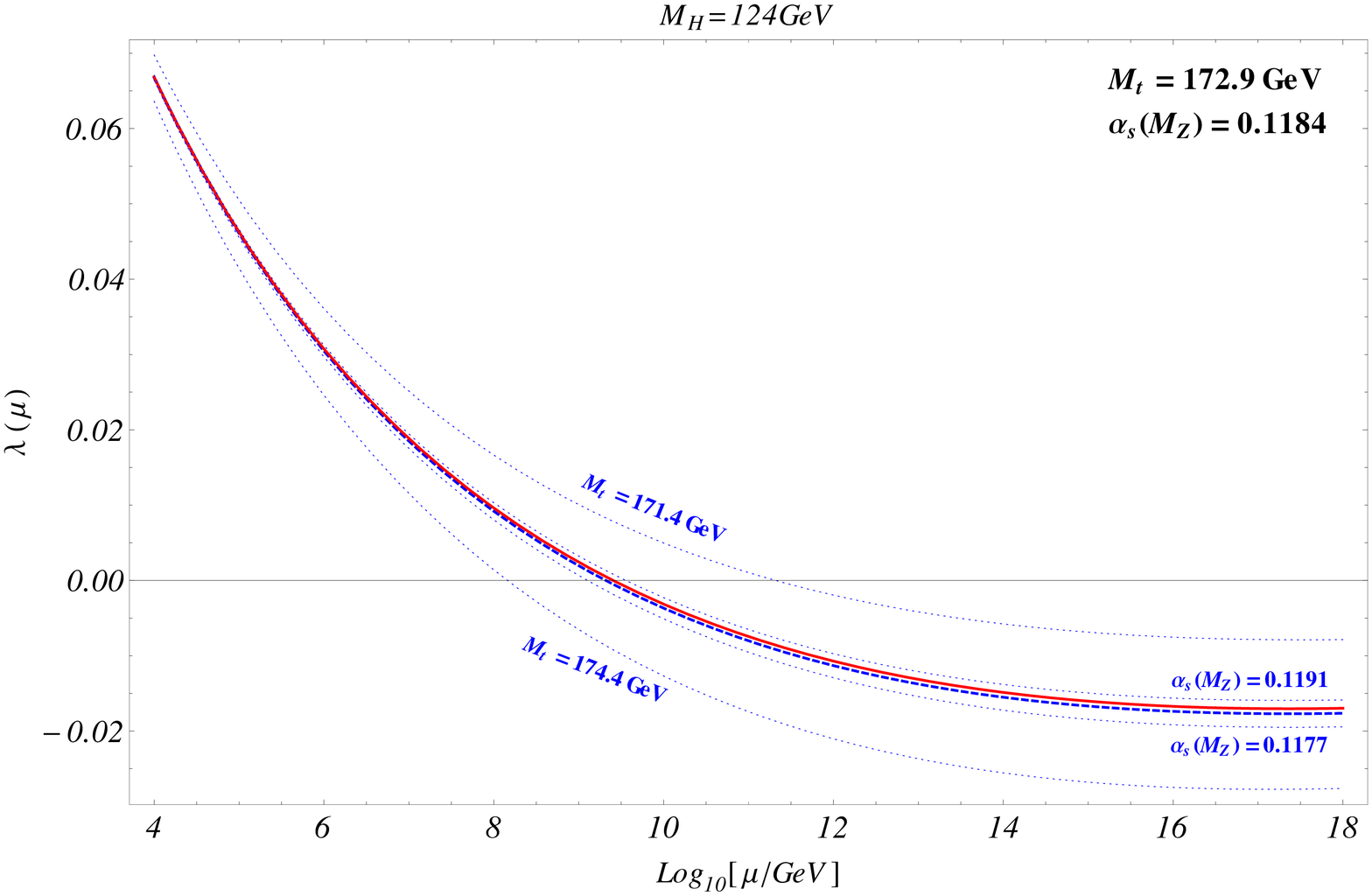} 
\includegraphics[width=0.95\textwidth]{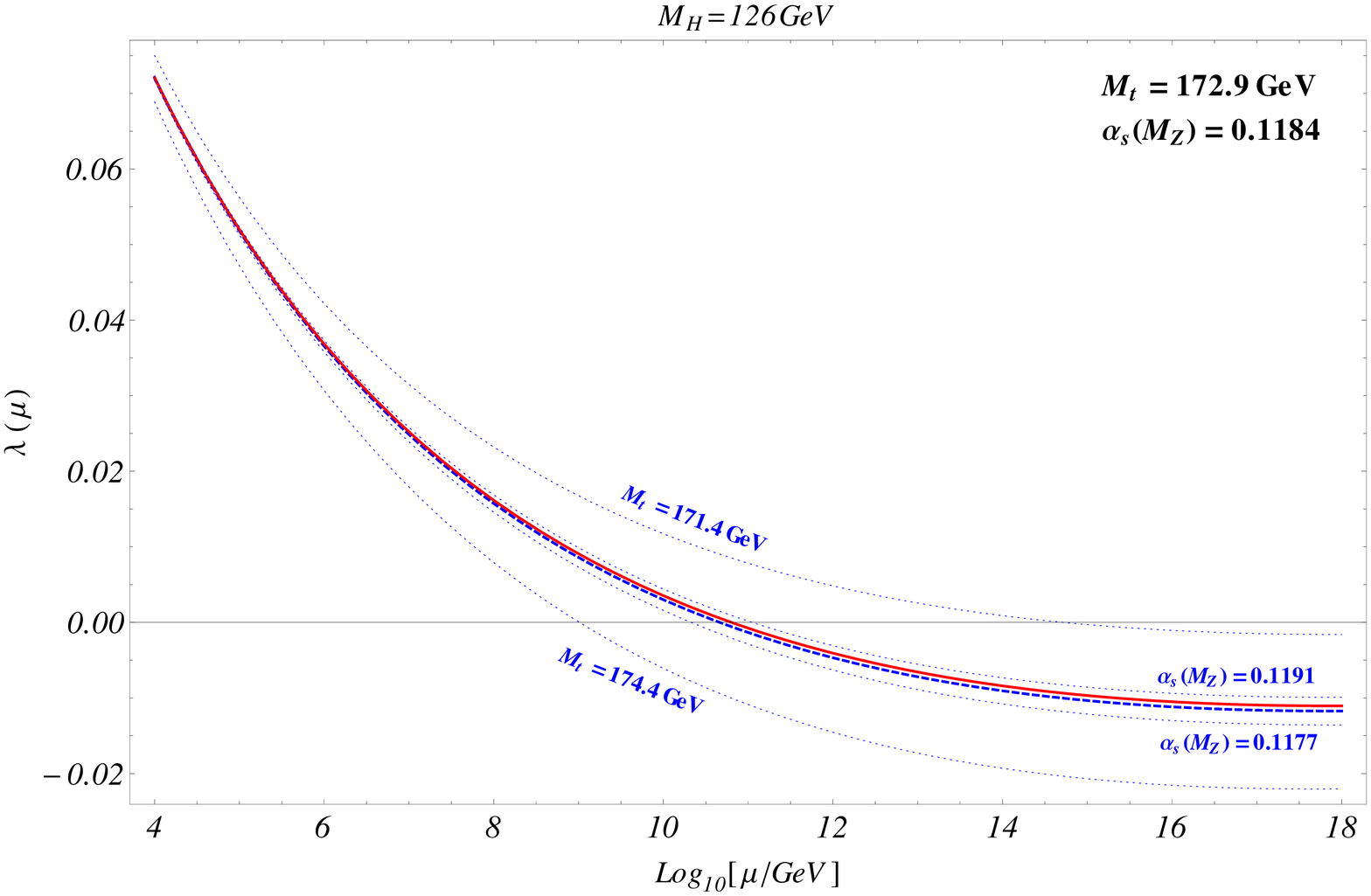} 
\caption{Evolution of $\lambda$ with the scale $\mu$: \Blue{2 loop (dashed, blue)} and \Red{3 loop (continuous, red)} results;
 Uncertainties with respect to the two-loop result: $\pm 1\sigma_{\als}$, $\pm 1\sigma_{M_t}$ (dotted)}
\label{lambda124126}
\end{figure} 
For this we also include the
electroweak contributions up to the two-loop level. 
For recent detailed discussions of the SM vacuum stability and its dependence on
key parameters like the Higgs and the top mass see for example \cite{Masina:2012tz,Holthausen:2011aa,Bezrukov:2012sa,Degrassi:2012ry,Chetyrkin:2012rz}.
We start the evolution of $\lambda$ at the scale of the top mass and go
up to the Planck scale at $10^{18}$ GeV.
To find the starting values in the $\overline{\text{MS}}$-scheme
we have to match the physical parameters, like e.g. pole masses, to their 
$\overline{\text{MS}}$ counterparts.\footnote{We take the electroweak corrections at one-loop
and the QCD ones at two-loop level from \cite{Espinosa:2007qp,Hempfling:1994ar,Sirlin1986389}.}  
These matching relations  depend on
the exact values of the $\overline{\text{MS}}$ parameter $\als(M_Z)$, the pole mass $M_{{t}}$ of the top quark
and of course the mass of the Higgs boson
$M_H$. We use \cite{pdg}
\be \begin{split}
 \als(M_Z)=0.1184 \pm 0.0007,\qquad
M_{{t}}=172.9 \pm 0.6 \pm 0.9 \text{ GeV}.
\end{split}
\label{alsmt}
\ee
Fig. \ref{lambda124126} shows the evolution of $\lambda$ for the two cases
\mbox{$M_H=124$ GeV} and \mbox{$M_H=126$ GeV.}. The dependence of the $\lambda$-running on the parameters $\als(M_Z)$ and $M_{{t}}$ 
can be estimated from the shifted curves where these parameters are changed
by $\pm \sigma$ as given in eq.~(\ref{alsmt}).
Note that there is a considerable difference between
\mbox{$M_H=124$ GeV} and \mbox{$M_H=126$ GeV} which
means that the evolution of $\lambda$ is very sensitive to the value
of the Higgs mass. Given a fixed value for $M_H$ the largest
uncertainty lies in the exact value of the top mass. The second
largest uncertainty comes from $\als$. The total effect due to the
three-loop part of the $\beta$-functions, an extension of the vacuum stability to larger scales, 
is a little smaller than the $\als$ uncertainty.
Note that an analysis including  higher order corrections for the matching between 
the pole masses and $\overline{\text{MS}}$-masses has been performed in \cite{Bezrukov:2012sa,EliasMiro:2011aa}.

In conclusion it can be said that for a Higgs mass in the vicinity of $125$ GeV the
question of vacuum stability cannot be resolved with certainty. Although
it looks as if $\lambda$ becomes indeed negative at high scales the experimental uncertainties, mainly on
the Higgs mass and the top mass, are too large at the moment. However, this is a very good motivation to attempt higher
precision experiments and calculations in the near future.

{\bf Acknowledgments}\\
I want to thank my collaborator on this project K. G. Chetyrkin for invaluable discussions
and J. K\"uhn for his support and useful comments. 
This work has been supported by the Deutsche Forschungsgemeinschaft in the
Sonderforschungsbereich/Transregio SFB/TR-9 ``Computational Particle
Physics'' and the Graduiertenkolleg ``Elementarteilchenphysik
bei h\"ochsten Energien und h\"ochster Pr\"azission''

\bibliographystyle{utphys}

\footnotesize

\bibliography{LiteraturSM}

\end{document}